\documentclass[epj-spec,final]{svjour}

\usepackage{graphicx}
\usepackage{dcolumn}
\usepackage{bm}

\usepackage{amsmath,amssymb}
\usepackage[numbers,sort&compress]{natbib}

\newcommand\diff{\mathrm{d}}

\begin{document}

\title{Localization phenomena in models of ion-conducting glass formers}


\author{J\"urgen Horbach\inst{1}\fnmsep\thanks{\email{juergen.horbach@dlr.de}}
\and Thomas Voigtmann\inst{1,2}
\and Felix H\"ofling\inst{3,4}
\and Thomas Franosch\inst{5,6}}
\institute{
Institut f\"ur Materialphysik im Weltraum, Deutsches Zentrum f\"ur Luft- und
Raumfahrt (DLR), 51170 K\"oln, Germany
\and
Zukunftskolleg und Fachbereich Physik, Universit\"at Konstanz, 78457 Konstanz, Germany
\and
Rudolf Peierls Centre for Theoretical Physics, 1~Keble Road, Oxford OX1 3NP, England,
United Kingdom
\and
Max-Planck-Institut f\"ur Metallforschung, Heisenbergstra{\ss}e 3, 70569
Stuttgart and Institut f\"ur Theore\-tische und Angewandte Physik, Universit\"at
Stuttgart, Pfaffenwaldring 57, 70569 Stuttgart, Germany
\and
Institut f\"ur Theoretische Physik, Universit\"at Erlangen-N\"urnberg,
Staudtstra{\ss}e 7,
91058 Erlangen, Germany
\and
Arnold Sommerfeld Center for Theoretical Physics (ASC) and Center for NanoScience (CeNS),
Fakult\"at f\"ur Physik, Ludwig-Maximilians-Universit\"at M\"unchen, Theresienstra{\ss}e 37,
80333 M\"unchen, Germany
}

\abstract{
The mass transport in soft-sphere mixtures of small and big particles
as well as in the disordered Lorentz gas (LG) model is studied using
molecular dynamics (MD) computer simulations. The soft-sphere mixture
shows anomalous small-particle diffusion signifying a localization
transition separate from the big-particle glass transition.
Switching off small-particle excluded volume
constraints slows down the small-particle dynamics, as indicated by
incoherent intermediate scattering functions. A comparison of logarithmic
time derivatives of the mean-squared displacements reveals qualitative
similarities between the localization transition in the soft-sphere
mixture and its counterpart in the LG.  Nevertheless, qualitative
differences emphasize the need for further research
elucidating the connection between both models.
}
\maketitle
\section{Introduction}
\label{intro}
Transport in heterogeneous disordered materials and in porous media
arises in many diverse subjects of science and engineering such as
water resource management, oil recovery, the physics of crowded
biological systems, glass technology or even the geophysical
understanding of the eruption of volcanos \cite{Dingwell:1996}.
Examples are fast ion transport in alkali-silicate melts and other
industrially relevant glass formers \cite{Bunde:1998}, protein motion
on cell membranes~\cite{Kusumi:2005,Avidin:2010} and within the living
cell~\cite{Weiss:2003}, among others.  Common to all these materials
is that they consist of at least two components, one of which (the
``fast component'') is responsible for the observed transport phenomena,
while the other forms the heterogeneous matrix and is either completely
frozen as, e.g., in porous media, or relaxes orders of magnitude slower,
as in glass-forming ion conductors.

A well-known reference point
for diffusion in random environments is the Lorentz gas
(LG)~\cite{Goetze:1981a,vanBeijeren:1982}: considering a single
particle diffusing through a set of fixed obstacles, one essentially
arrives at an off-lattice model for the transport through porous media or
ion-conducting systems. The LG exhibits a localization transition where
diffusion of the tracer ceases altogether if the density of scatterers
exceeds a certain value; this transition is understood as a dynamic
critical phenomenon connected with the percolation of void space in the
heterogeneous background medium~\cite{Lorentz_PRL:2006,Lorentz_JCP:2008}.

Obviously, several aspects of ion-conducting systems are described in
an oversimplified manner by the LG model.  In a real ion conductor,
the frozen environment is not formed by fixed obstacles that are
randomly distributed. Instead, the ``obstacle particles'' exhibit
structural correlations (at least on short length scales) and
they are subject to thermal motion.  Moreover, in most ion-conducting
systems the mobile ions cannot be considered as non-interacting tracer
particles and the correlations between ions have to be taken into
account. Realistic modeling
for sodium silica melts~\cite{Horbach:2002,Meyer:2004,Voigtmann:2006}
reveals that the silica network can constitute such a quasi-arrested
array of correlated obstacles through which sodium ions meander
on preferential diffusion pathways. The chemical properties of the
silica melt to form a tetrahedral network appears not to be
essential: a similar decoupling of diffusive transport is
observed in dense size-asymmetric Yukawa melts~\cite{Kikuchi:2007}
and binary soft-sphere mixtures~\cite{Moreno:2006+Moreno:2006a}.
Mode-coupling theory has been used to predict it for size-disparate hard-sphere
systems \cite{Bosse:1987,Bosse:1995}.
As shown recently by
molecular dynamics (MD) computer simulations~\cite{Voigtmann:2009},
such systems may exhibit a glass transition where
the fast component remains mobile even for high-density states at which
the matrix does not show relaxation over the entire simulation time
window. At these high densities, evidence was found that the small
particles show anomalous diffusion and approach a localization transition
that bears resemblance to the LG transition. Similar findings hold for
an entirely
frozen matrix, where non-trivial predictions of a mode-coupling theory
developed by
Krakoviack~\cite{Krakoviack:2005+Krakoviack:2007+Krakoviack:2009}
have been tested
recently~\cite{Kurzidim:2009,Kim:2009}. However, it is an open question to
what extent the small-particle localization dynamics in binary soft-sphere
mixtures can be understood as a dynamic critical phenomenon that falls
into the same universality class as the localization transition of
the LG~\cite{Lorentz_PRL:2006,Lorentz_JCP:2008} and other continuum
percolation models~\cite{benAvraham:DiffusionInFractals}.  In particular,
the question arises how the localization of the small particles is
affected by interactions between the mobile tracers.

In this contribution, we present first steps to address the latter
issues.  We compare on a qualitative level the anomalous diffusion of a
binary soft-sphere mixture with that of a LG. The effect of collective
interactions among the small particles is investigated by switching off
their interactions.

\section{Results}
\begin{figure}
\sidecaption
\includegraphics[width=0.6\linewidth]{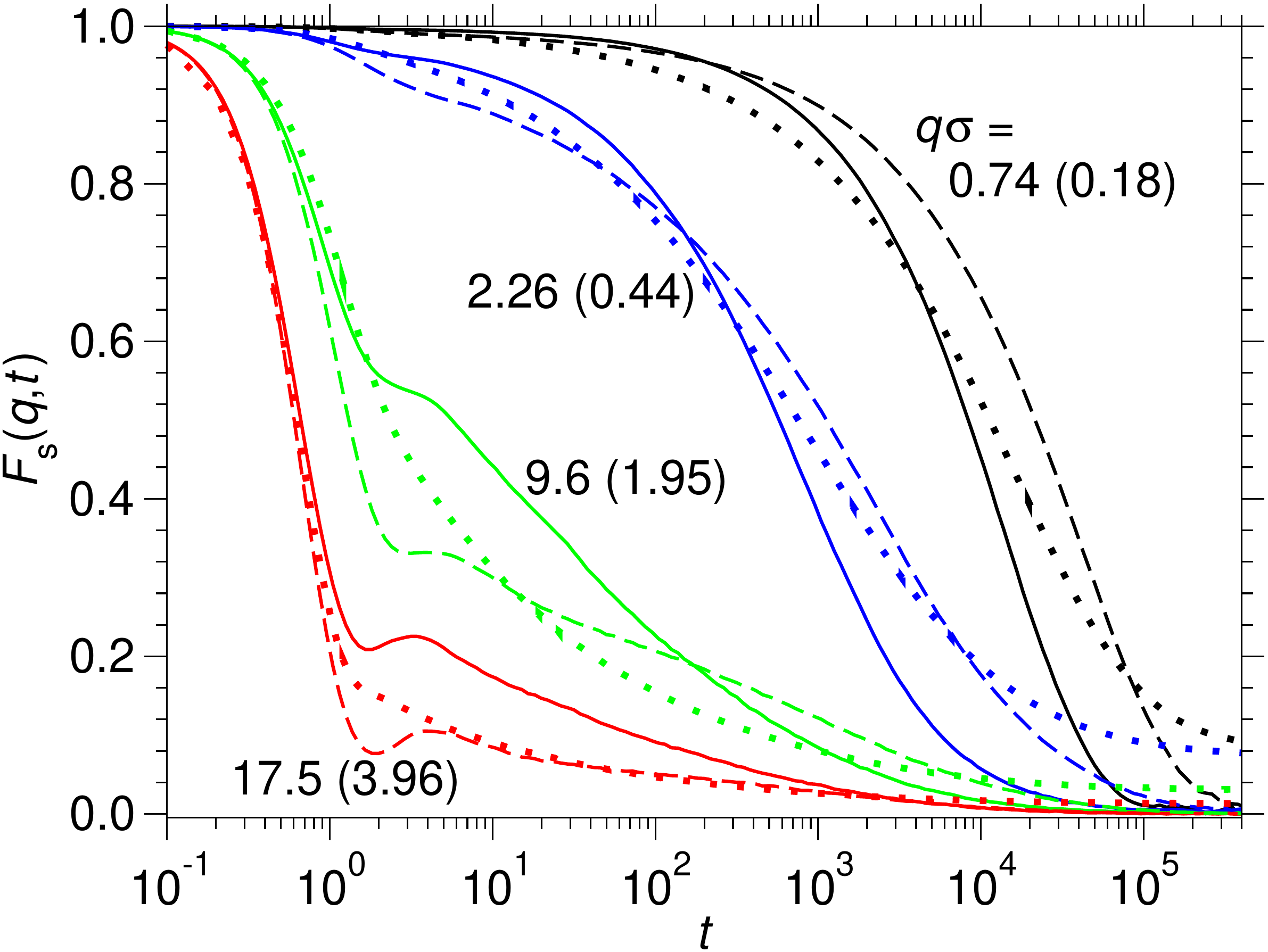}
\caption{Incoherent intermediate scattering function $F_{\rm s}(q,t)$
for the small/tracer particles.
Solid and dashed lines correspond respectively to the BSSM with
and without interactions among the small particles, while dotted lines
represent the LG model. For the BSSM, the wavenumbers are in the range
$0.74\le q \sigma \le 17.5$, as indicated (with $\sigma$ the diameter of
the big particles). The wavenumbers in brackets correspond to the LG.
The number density of the BSSM is $\rho=3.26$.
For the LG, the packing fraction
of the obstacles is $n^*=0.75$.
\label{fig_bm}}
\end{figure}
We performed MD simulations of an equimolar binary mixture of purely
repulsive soft spheres (BSSM) with a size ratio of 0.35. Diameters are
chosen additively, and nonadditive energetic interactions further decouple
the species.  Temperature is unity, and all masses are equal.
For details, see Ref.~\citealp{Voigtmann:2009}.
MD simulations of the LG were carried
out using an event-driven algorithm and randomly distributed,
possibly overlapping hard sphere obstacles.
The dimensionless obstacle
density $n^*$ is the only control parameter of the model; details
can be found in Ref.~\citealp{Lorentz_JCP:2008}.

Figure~\ref{fig_bm} shows the time dependence of the incoherent
intermediate scattering function, $F_{\rm s}(q,t)$, of the small particles
for different values of the wavenumber $q$, as indicated.  For the BSSM,
at a number density $\rho=3.26$, the big particles form a glass
and the transport of the small particles is characterized by
anomalous subdiffusive motion crossing over to ordinary diffusion at
long times.
Two kinds of BSSM systems are considered in Fig.~\ref{fig_bm}:
the solid lines correspond to the system with interacting small particles,
while the dashed lines correspond to a system where the small--small
interaction potential is set to zero, keeping all other
parameters fixed. Structure and dynamics of the big particles
as measured through the partial static structure factor
are unaffected by this \cite{Voigtmann:2009}.  Surprisingly, at fixed $q$ the
curves for the fully interacting case decay more rapidly than those for the
non-interacting case. In Ref.~\citealp{Voigtmann:2009}, we have proposed
that this difference arises because small-particle interactions result in
a more directional and
thus more effective diffusive motion.

Also shown in Fig.~\ref{fig_bm} are results for the LG at a packing
fraction of the obstacles $n^*=0.75$.  At this value of $n^*$, the
decay of $F_{\rm s}(q,t)$ takes place on a similar time scale provided
that one reduces the wavenumbers by about a factor of 4 (as done in
Fig.~\ref{fig_bm}) to match the length scales of the LG and the BSSM
systems. The curves for the LG demonstrate the absence of a ``glassy''
plateau at intermediate times. At long times, they decay to a plateau of
different origin: the presence of
closed, finite pockets in the obstacle matrix leads to a localization of tracer
particles. However, apart from that the decay of the LG correlators is
similar to that for the BSSM systems.  So the question arises whether
the transport as seen for the non-interacting particles in the binary
mixture is similar to that seen in the LG.

\begin{figure}\sidecaption
\includegraphics[width=0.5\linewidth]{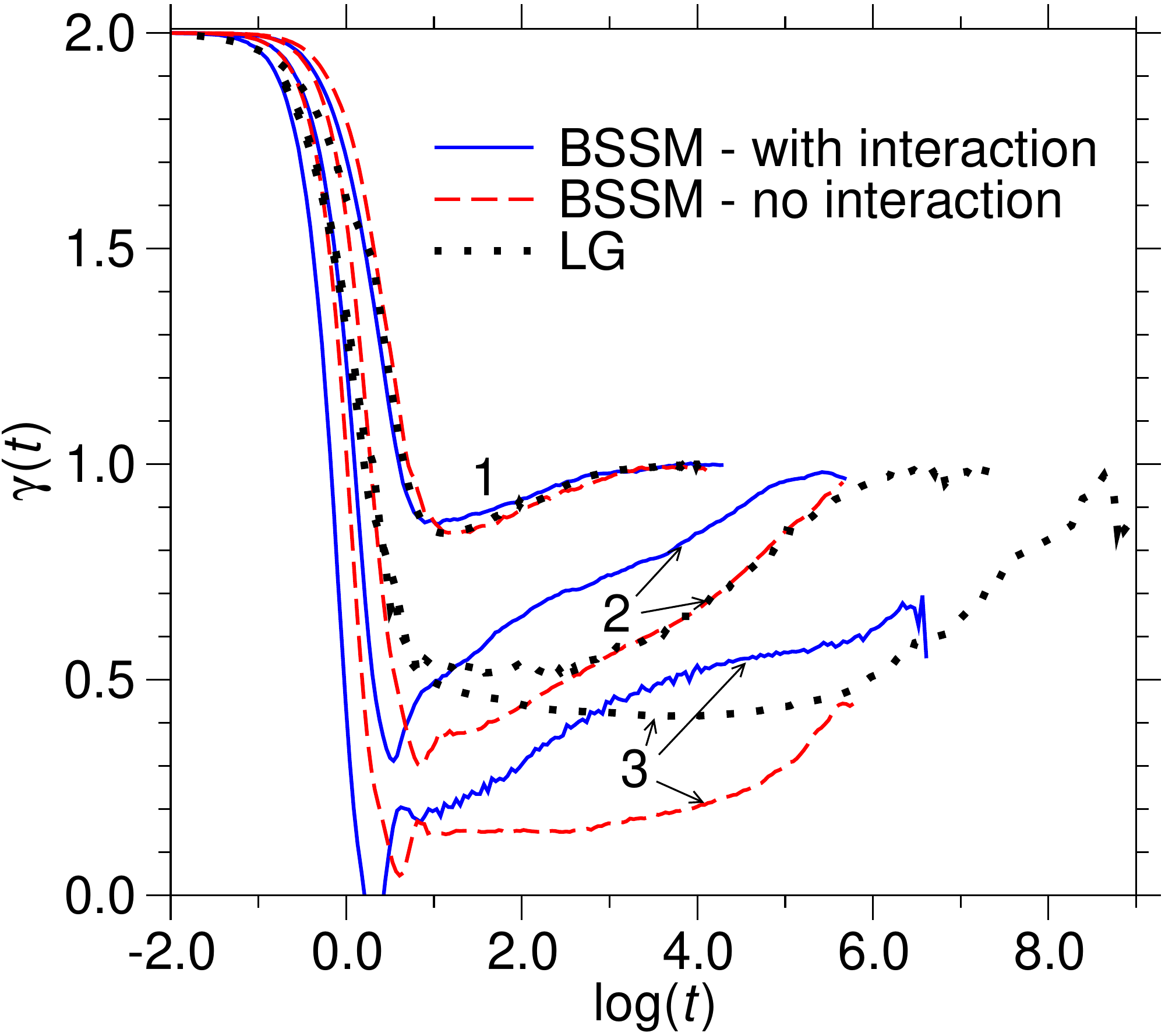}
\caption{
Effective exponent, as obtained from the logarithmic derivative
$\gamma(t)$ of the mobile particles for the binary mixture with and
without interactions between the small particles as well as for the
LG. Results are shown for different densities: the three sets marked with
number 1 correspond to $\rho=2.296$ and $n^*=0.40$, those with number 2
to $\rho=3.257$ and $n^*=0.75$ and those with number 3 to $\rho=4.215$
and $n^*=0.82$ (here, $\rho$ and $n^*$ are the densities of the binary
mixtures and the LG, respectively). \label{newfig1}}
\end{figure}
To address this question, we compare the logarithmic derivative
$\gamma(t)=\diff[\log\delta r^2(t)]/\diff(\log t)$ for the two binary
mixtures and for the LG in Fig.~\ref{newfig1}; $\gamma(t)$ is calculated
from the mean-squared displacements $\delta r^2(t)$ of the mobile
particles in both systems.  The quantity
$\gamma(t)$ has the meaning of an effective exponent that crosses over
from $\gamma(0)=2$ (ballistic short-time motion) to $\gamma(\infty)=1$
for diffusive particles in the liquid.  For localized particles,
$\gamma(\infty)=0$ is expected.

The selected obstacle densities for the LG in Fig.~\ref{newfig1},
$n^*=0.40$, 0.75 and 0.82, are below the critical density of the
localization transition, $n_c^* = 0.837$~\cite{Lorentz_JCP:2008}.
At criticality $n^*=n^*_c$, the exponent $\gamma(t)\approx 0.32$ is approached
at long times~\cite{Lorentz_PRL:2006}, thus confirming
the universal subdiffusive behavior expected from renormalization group
arguments (refer to Ref.~\citealp{Lorentz_JCP:2008} for a detailed discussion).
Below $n_c^*$, such anomalous diffusion is
seen in a finite, intermediate time window and universal corrections to
scaling~\cite{Percolation_EPL:2008} lead to the observation of an
effective exponent that is larger than the universal critical exponent.
At $n^*=0.82$ for example, an effective exponent $\gamma\approx
0.4$ is observed over about four orders of magnitude in time, see
Fig.~\ref{newfig1}.

At low densities (set of curves ``1'' in Fig.~\ref{newfig1}), the binary
mixtures with and without interactions between the small particles show almost
identical behavior as the LG with respect to $\gamma(t)$.
Nevertheless, there are quantitative differences: At higher densities
the effective exponent of the BSSM without interaction becomes rather small,
the corresponding
mean-square displacement remains of the order of the interparticle
distance, and the cage effect interferes strongly with the regime of
anomalous
diffusion. Turning on the interaction, a window of subdiffusion emerges,
yet the effective exponent drifts gradually from 0.5 to 0.6.
In contrast, the  LG displays clear subdiffusive behavior  over several
orders of magnitude in time  as manifested in an almost constant
$\gamma(t)$, as mentioned above.

Clearly, a simplistic mapping between the BSSM and the overlapping LG is not
obvious and may not exist due to the highly simplified character of the LG.
Thus, the investigation of LG variants with increasing complexity is desirable,
which close the gap to the BSSM. For example, a better approximation of the
matrix structure would be obtained by introducing correlated obstacles in the
LG (``non-overlapping LG''). Another issue is raised by the hard interaction
potentials in the LG.  While a tracer with constant velocity between soft
obstacles is easily mapped to the hard sphere model, the percolation transition
is already smeared out for thermal tracers (or tracers in contact with a heat
bath) due to the average over the Maxwell distribution. All these issues are
the subject of forthcoming studies.


This collaborative work has been started as part of the Research Unit
FOR~1394 of Deutsche Forschungsgemeinschaft (DFG).
We acknowledge a substantial grant of computer time at the NIC J\"ulich.
Th.~V.\ acknowledges funding from the Helmholtz-Gemeinschaft (VH-NG~406)
and the Zukunftskolleg Konstanz.




\end{document}